\documentclass[3p,times,procedia]{elsarticle}
\usepackage{nupha_ecrc}

\usepackage[utf8]{inputenc}

\usepackage{mathtools}
\usepackage{amsfonts}
\usepackage{mathrsfs}
\usepackage{slashed}

\usepackage{graphicx}
\usepackage{subcaption}
\usepackage[dvipsnames]{xcolor}
\usepackage{array}

\usepackage{placeins}

\usepackage{xspace}
\usepackage{hyperref}
\usepackage[symbol]{footmisc}

\usepackage{booktabs}

\hypersetup{
	colorlinks,
	linkcolor={red!75!black},
	citecolor={blue!75!black},
	urlcolor={blue!75!black},
	pdftitle={Time-evolution of fluctuations as signal of the phase transition dynamics in a QCD-assisted transport approach},
	pdfauthor={Bluhm, Jiang, Nahrgang,Pawlowski, Rennecke, Wink},
	pdfkeywords={Spectral function} {Analytic continuation}
	{Correlations functions} {Transport},
	bookmarksopen=true,
	bookmarksopenlevel=2,
	bookmarksnumbered=true
}

\volume{00}

\firstpage{1}

\journalname{Nuclear Physics A}

\runauth{M. Bluhm, Y. Jiang, M. Nahrgang, J. M. Pawlowski, F. Rennecke, N. Wink}

\jid{nupha}

\jnltitlelogo{Nuclear Physics A}

\usepackage{amssymb}

\usepackage[figuresright]{rotating}

\begin{document}

\begin{frontmatter}



\dochead{XXVIIth International Conference on Ultrarelativistic Nucleus-Nucleus Collisions\\ (Quark Matter 2018)}

\title{Time-evolution of fluctuations as signal of the phase transition dynamics in a QCD-assisted transport approach}


\author[lbl:breslau]{M.~Bluhm}
\author[lbl:beihang]{\hspace{-0.05cm}Y.~Jiang}
\author[lbl:nantes,lbl:emmi]{\hspace{-0.05cm}M.~Nahrgang}
\author[lbl:emmi,lbl:hd]{\hspace{-0.05cm}J.M.~Pawlowski}
\author[lbl:bnl]{\hspace{-0.05cm}F.~Rennecke}
\author[lbl:hd]{\hspace{-0.05cm}N.~Wink\footnote[1]{\vspace{-3cm}Speaker}}

\address[lbl:breslau]{Institute of Theoretical Physics, University of Wroclaw, PL-50204 Wroclaw, Poland}
\address[lbl:beihang]{School of Physics and Nuclear Energy Engineering, Beihang University, Beijing 100191, China}
\address[lbl:nantes]{SUBATECH UMR 6457 (IMT Atlantique, Universit\'e de Nantes, IN2P3/CNRS), 4 rue Alfred Kastler, 44307 Nantes, France}
\address[lbl:emmi]{ExtreMe Matter Institute EMMI, GSI, Planckstraße 1, D-64291 Darmstadt, Germany}
\address[lbl:hd]{Institut f\"ur Theoretische Physik, Universit\"at Heidelberg, Philosophenweg 16, D-69120 Heidelberg, Germany}
\address[lbl:bnl]{Physics Department, Brookhaven National Laboratory, Upton, NY 11973, USA}

\begin{abstract}
For the understanding of fluctuation measurements in heavy-ion collisions 
it is crucial to develop quantitatively reliable dynamical descriptions which take the 
non-perturbative nature of QCD near the phase transition into account. We 
discuss a novel QCD-assisted transport approach based on 
non-equilibrium chiral fluid dynamics and the effective action of low energy 
QCD. In this framework, we study the time-evolution of fluctuation measures of 
the critical mode, notably the kurtosis, for a non-expanding system. From this, we can estimate 
the equilibration times of critical mode fluctuations in the QCD phase diagram. These allow 
us to identify both the phase boundary and the critical region near the QCD critical point.
\end{abstract}


\end{frontmatter}


\section{Introduction}
\label{sec:Introduction}
In the last decades tremendous progress has been made in understanding
the phase structure of strongly interacting matter. This has been
driven by heavy-ion collisions on the experimental side and by Lattice
QCD, functional approaches to QCD, perturbation theory and effective
theories on the theoretical side. Most notably the existence of a
deconfined phase, i.e. the Quark-Gluon plasma, and its phase
transition at vanishing and small net-baryon density are by now well
established from both theory and experiment. Despite these efforts,
the situation at larger densities is less clear. In order to verify
the existence of a possible critical endpoint in the phase diagram of
QCD, a more detailed understanding of the connection between the
equilibrium phase structure and the highly dynamical non-equilibrium
situation created in heavy-ion collisions needs to be
established. Only then firm conclusions can unambiguously be drawn
from fluctuation measurements.

First attempts in this direction based on non-equilibrium chiral fluid
dynamics studies connected with an effective mean-field model for QCD
\cite{Nahrgang:2011mg,Nahrgang:2011mv,Herold:2016uvv}. A recent
approach \cite{Nahrgang:2018afz} studies a fully interacting
stochastic description of the non-equilibrium evolution of fluctuation
observables.  In the present work, we explore a novel method which is
capable of connecting the equilibrium physics of QCD, obtained beyond
mean field within the Functional Renormalization Group (FRG) approach
to QCD, with the non-equilibrium evolution around an equilibrium state
in a systematic manner. In a first study, we apply this method to the
time-evolution of the critical mode around the various equilibrium
states in the phase diagram of a 2+1 flavour Quark-Meson model. This
is achieved by solving a transport equation with the corresponding
linear response functions of a given equilibrium state as input.

In \autoref{sec:Equilibrium}, we describe the calculation of the
necessary equilibrium input, i.e. the effective potential and the
spectral functions of the sigma meson. In \autoref{sec:Transport}, the
dynamical evolution of the critical mode around this equilibrium
result is described.

\section{Equilibrium linear response functions}
\label{sec:Equilibrium}

The calculation of equilibrium correlation functions needed as input
for the transport evolution utilizes the Functional Renormalization
Group. The FRG is a versatile, first principle tool that has been
applied successfully to QCD, see e.g.~\cite{Cyrol:2017ewj}, and
low-energy effective versions thereof, see e.g.~\cite{Herbst:2010rf,
  Rennecke:2016tkm}. The advantage of the FRG in the present context
is that it allows for the computation of the phase structure, i.e. the
effective potential, and momentum dependent correlation functions
within a unified framework.

The equilibrium part of our work, i.e.~the equation of state and the
equilibrium correlation functions, is based on a 2+1 flavour study of
a low-energy effective description of QCD, where the dynamics of
constituent quarks as well as the lowest scalar- and pseudoscalar
meson nonets are taken into account \cite{Rennecke:2016tkm}. It
captures, by design, the relevant physical effects at small chemical
potential $\mu$ and temperatures \mbox{$T \lesssim
  T_c$}. Additionally, it features a critical endpoint which is in the
same static universality class as the one potentially present in
QCD. Therefore this model provides a well-suited base for studying how
dynamical non-equilibrium effects manifest themselves in observables.

\begin{figure}[t]
	\centering
	\includegraphics[width=0.5\textwidth]{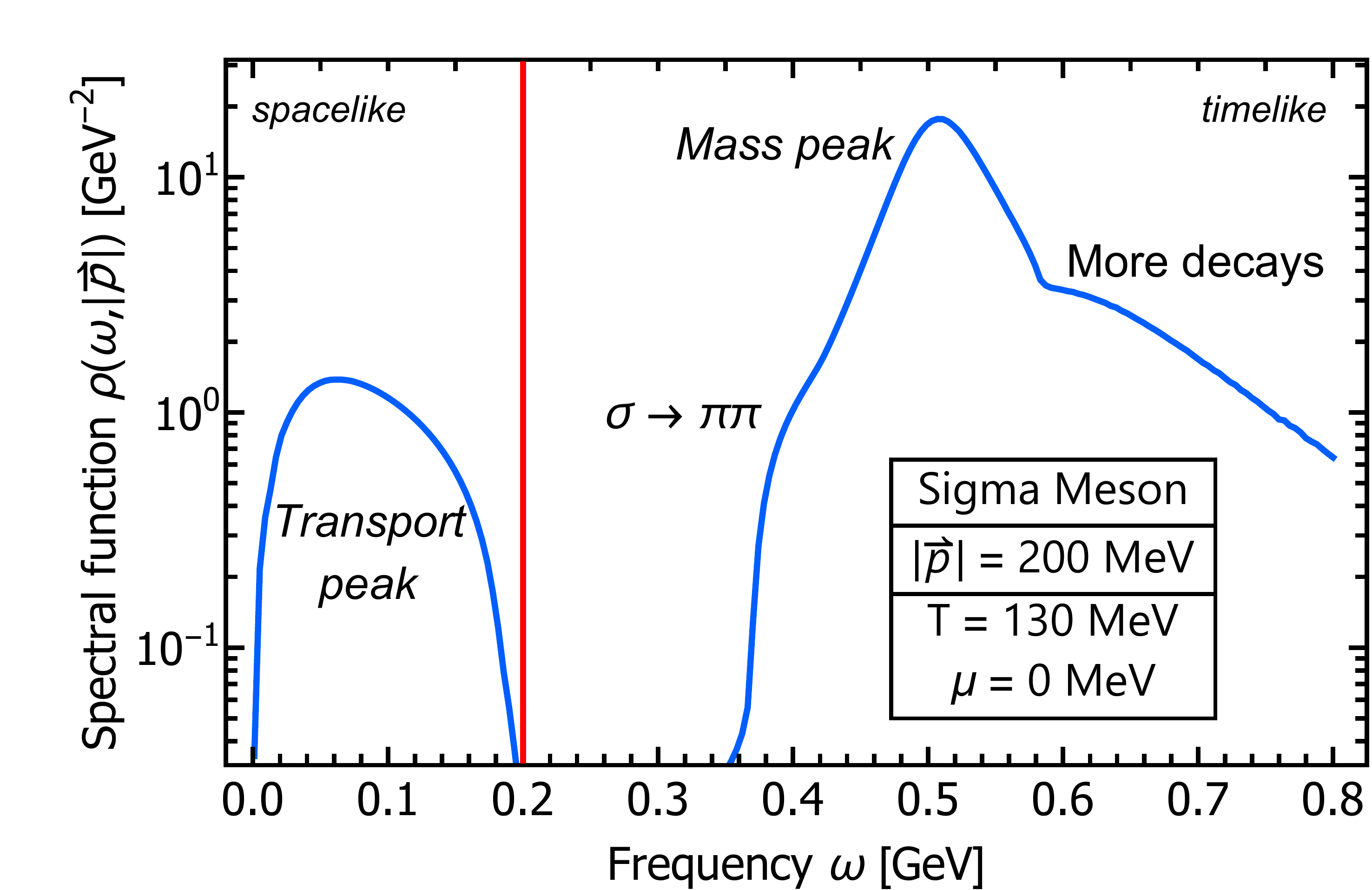}
	\caption{ Spectral function of the sigma meson at $T=130$~MeV,
          $\mu=0$~MeV in the phase diagram. The transport peak and the
          mass peak are associated with the diffusion in the transport
          equation. A detailed discussion of the seen structures can
          be found in e.g.~\cite{Pawlowski:2017gxj}.  }
	\label{fig:sigma_spectral}
\end{figure}

In general, spectral functions can be obtained either via analytically
continuing numerical data, see e.g.~\cite{Cyrol:2018xeq} or via a
direct computation from analytically continued equations, see
e.g.~\cite{Floerchinger:2011sc,Kamikado:2013sia}. If possible, the
latter is preferred and also the option utilized in this work. The
spectral functions of the sigma meson are calculated similarly to
\cite{Tripolt:2013jra,Pawlowski:2017gxj} with suitable modifications
in order to take non-trivial wave-function renormalizations into
account. As a result we have access to the two-point correlator
$\Gamma_{\sigma\sigma}^{(2)}(\omega,|\vec{p}|)$, depending on an
external frequency $\omega$ and an external momentum $\vec{p}$, as
well as momentum independent vertices $\Gamma_{\sigma^n}^{(n)}$ which
are extracted from the full effective potential computed in
\cite{Rennecke:2016tkm}. An exemplary spectral function is shown in
\autoref{fig:sigma_spectral}. The two main features that influence the
behaviour of the dynamical evolution are the transport peak and the
mass peak. The transport peak, if present at small frequencies
$\omega<|\vec{p}|$, dominates the long range behaviour of the sigma
field. The mass peak, instead, becomes the driving force for the
evolution dynamics when the transport peak is absent, e.g.\ in the
vacuum.

\section{Time-evolution of fluctuation measures}
\label{sec:Transport}
We are now in the position to study the time-evolution of the critical
mode and its event-by-event fluctuations. For this purpose, we solve
the Langevin-type transport equation
\begin{align}
\label{equ:Langevin}
	\frac{\mathrm{d}\Gamma}{\mathrm{d}\sigma} = \xi \,.
\end{align}
In \eqref{equ:Langevin}, the equation of motion contains a kinetic
term related to the real part of $\Gamma^{\, (2)}_{\sigma\sigma}$, a
diffusion term sensitive to the imaginary part of
$\Gamma^{\, (2)}_{\sigma\sigma}$, and the effective potential,
discussed in \autoref{sec:Equilibrium}, while $\xi$ represents the
noise field chosen such that the fluctuation-dissipation balance is
guaranteed.

For the numerical results presented in the following we consider the
critical mode to be spatially isotropic,
i.e.~$\sigma(\vec{x},t)=\sigma(r,t)$, where we split
$\sigma=\sigma_0+\delta\sigma$. We study the time-evolution of the
critical fluctuations for a system subject to a sudden quench from
high temperatures to a specific point in the QCD phase
diagram. Accordingly, the system is initialized such that
$\sigma(r,t=0)=0$ and $\partial_t\sigma(r,t=0)=0$ which implies that
the initial fluctuations $\delta\sigma(t=0)$ are of the magnitude of
the equilibrium value $\sigma_0$ after the quench. Moreover, we
consider spatially constant Gaussian white noise, with zero mean and a
variance given as~\cite{Nahrgang:2011mg}
\begin{equation}
\label{equ:noisevariance}
 \langle\xi(t)\xi(t')\rangle=\frac1V\delta(t-t')m_\sigma\eta\coth\left(\frac{m_\sigma}{2T}\right)
 \, ,
\end{equation}
where the diffusion coefficient $\eta$ is extracted from the imaginary
part of $\Gamma^{\, (2)}_{\sigma\sigma}$, such that the
fluctuation-dissipation theorem is recovered.

\begin{figure}
	\begin{subfigure}[t]{0.47\textwidth}
	\includegraphics[width=\textwidth]{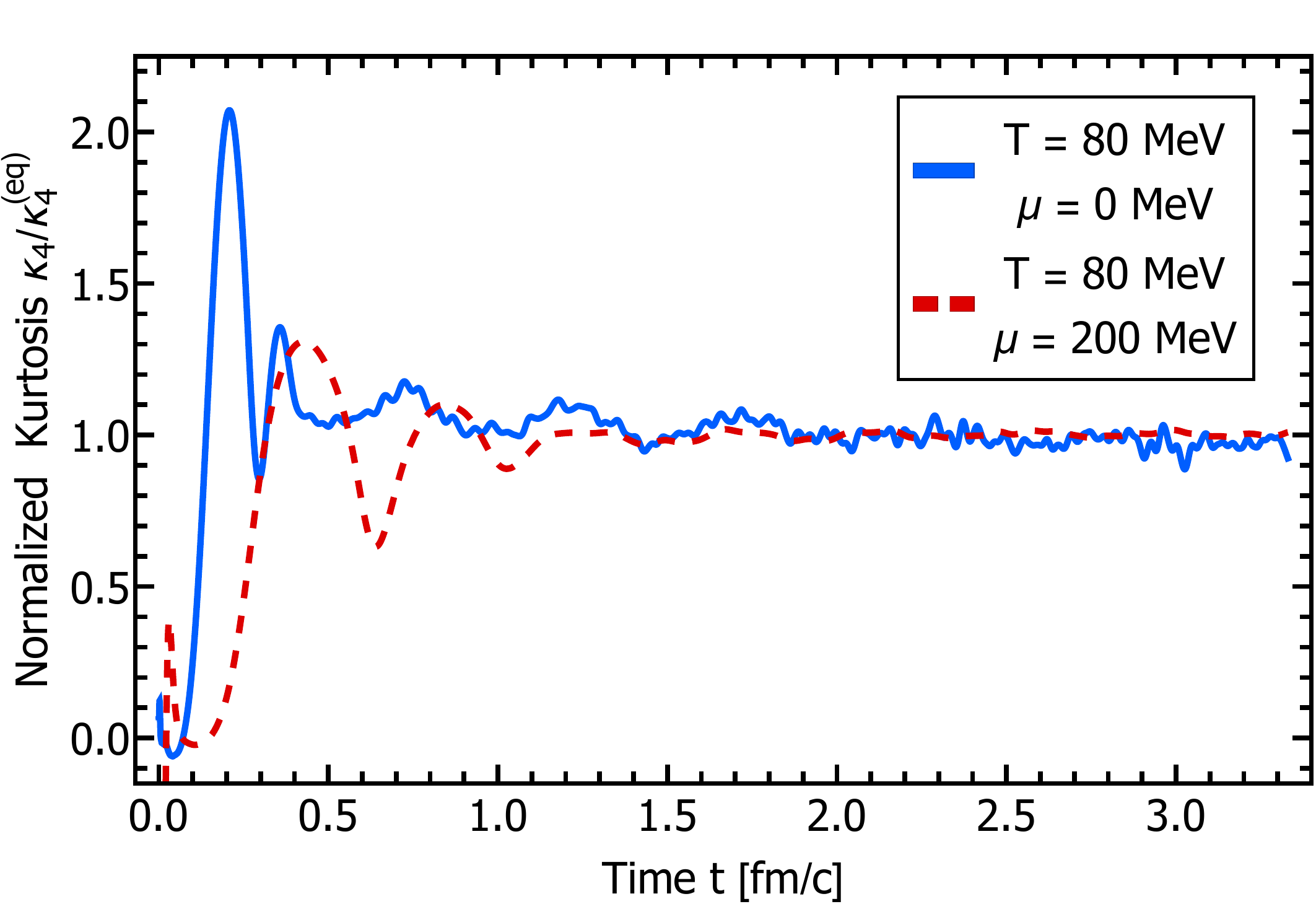}
	\end{subfigure}
	\hfill
	\begin{subfigure}[t]{0.47\textwidth}
	\includegraphics[width=\textwidth]{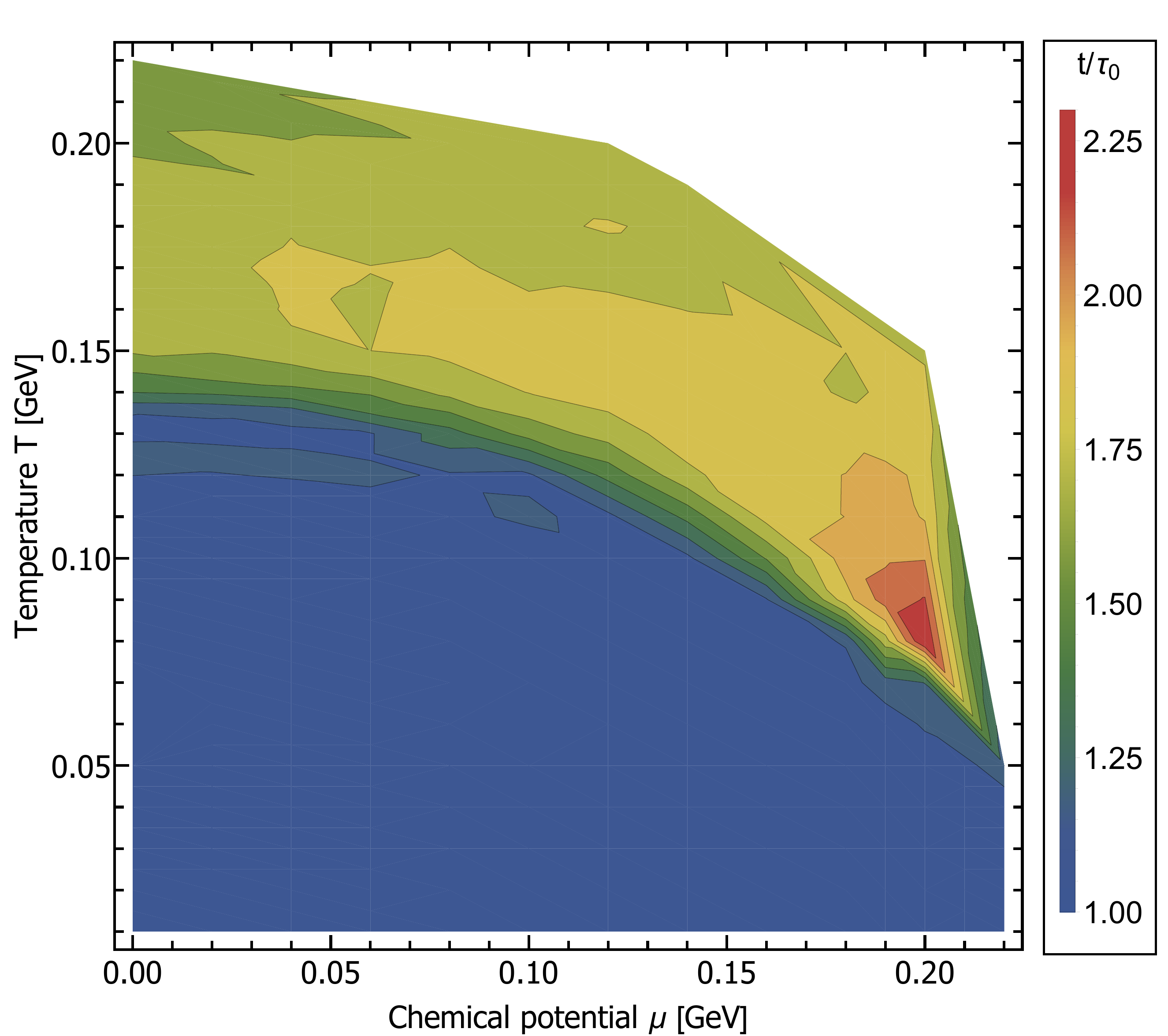}
	\end{subfigure}
	\caption{Left: Scaled kurtosis as a function of time for a
          quench from high $T$ to two different points in the phase
          diagram. Within statistical deviations, the equilibration
          time is found to be significantly increased near the
          critical endpoint (red, dashed curve) compared to a quench
          far away from it (blue, solid curve). Right: Equilibration
          time $t$ in units of $\tau_0\simeq 0.4$~fm/c in the QCD
          phase diagram based on the analysis of the scaled kurtosis
          in the quench scenario (see left panel).}
        \label{fig:equilibration}
\end{figure}
In \autoref{fig:equilibration} (left panel), we show, as an example,
the time-evolution of the kurtosis scaled by its late-time equilibrium
limit for the quench to two different points in the phase diagram. Far
away from the critical endpoint the scaled kurtosis exhibits a rather
quick equilibration while close to it the corresponding time scale is
clearly increased. For the quench through the phase boundary we
furthermore observe that the equilibrium value is approached from
above as the equilibrium kurtosis is larger near the phase boundary
than in the low-temperature phase.

Based on our preliminary results for the scaled kurtosis in the quench
scenario, we may estimate the equilibration time of the critical
fluctuations within the QCD phase diagram. This is shown in
\autoref{fig:equilibration} (right panel). One can clearly identify
both the phase boundary and the region near the critical endpoint and
observe the expected increase of the equilibration time in that
region. Nevertheless, we find this increase to be rather moderate
suggesting that phenomena associated with critical slowing down are
only moderately pronounced. This hints towards equilibrium dominated
measurements and, thus, to the feasibility of studying the QCD phase
diagram by means of heavy-ion collisions.

\section{Conclusion}
\label{sec:Conclusion}

In this work we developed a novel method to study dynamical
non-equilibrium effects in the phase diagram of the 2+1 flavour
Quark-Meson model. This method is valid not only in the scaling region
but across the entire region of the phase transition in QCD because it
contains critical and non-critical contributions to fluctuation
observables in a single framework. It builds upon the dynamics of a
quantum field around an equilibrium state with an appropriate
Langevin-type transport equation. We study the time-evolution of the
critical mode from a quench at high temperatures to different points
within the phase diagram. The time-evolution of different cumulants of
the sigma field was calculated, and an equilibration time was
extracted form the kurtosis. We find a moderate increase of the
equilibration time near the phase boundary and when approaching the
critical point. A region of critical slowing down is clearly
identifiable, but even close to the critical point only an enhancement
of roughly a factor of two is found.

In order to connect even closer to the full dynamics of a heavy-ion
collision, the scenario of a temperature quench will be subsequently
improved towards the full dynamics of the underlying quantum field
theory.




\section*{Acknowledgments}
The authors thank C.~Herold, V.~Koch and K.~Redlich for discussions.
The work of M.B.~is funded by the European Union’s Horizon 2020
research and innovation program under the Marie Sk\l{}odowska Curie
grant agreement No 665778 via the National Science Center, Poland,
under grant Polonez UMO-2016/21/P/ST2/04035. M.N.acknowledges the
support of the program ``Etoiles montantes en Pays de la Loire
2017''. F.R.~is supported by the DFG through grant \mbox{RE
  4174/1-1}. This work is supported by the ExtreMe Matter Institute
(EMMI) and the grant BMBF 05P12VHCTG.  It is part of and supported by
the DFG Collaborative Research Centre "SFB 1225 (ISOQUANT)".

\bibliographystyle{elsarticle-num}
\bibliography{transport_bib}

\begin{thebibliography}{10}
\expandafter\ifx\csname url\endcsname\relax
  \def\url#1{\texttt{#1}}\fi
\expandafter\ifx\csname urlprefix\endcsname\relax\def\urlprefix{URL }\fi
\expandafter\ifx\csname href\endcsname\relax
  \def\href#1#2{#2} \def\path#1{#1}\fi

\bibitem{Nahrgang:2011mg}
M.~Nahrgang, S.~Leupold, C.~Herold, M.~Bleicher, {Nonequilibrium chiral fluid
  dynamics including dissipation and noise}, Phys. Rev. C84 (2011) 024912.
\newblock \href {http://arxiv.org/abs/1105.0622} {\path{arXiv:1105.0622}},
  \href {http://dx.doi.org/10.1103/PhysRevC.84.024912}
  {\path{doi:10.1103/PhysRevC.84.024912}}.

\bibitem{Nahrgang:2011mv}
M.~Nahrgang, S.~Leupold, M.~Bleicher, {Equilibration and relaxation times at
  the chiral phase transition including reheating}, Phys. Lett. B711 (2012)
  109--116.
\newblock \href {http://arxiv.org/abs/1105.1396} {\path{arXiv:1105.1396}},
  \href {http://dx.doi.org/10.1016/j.physletb.2012.03.059}
  {\path{doi:10.1016/j.physletb.2012.03.059}}.

\bibitem{Herold:2016uvv}
C.~Herold, M.~Nahrgang, Y.~Yan, C.~Kobdaj, {Dynamical net-proton fluctuations
  near a QCD critical point}, Phys. Rev. C93~(2) (2016) 021902.
\newblock \href {http://arxiv.org/abs/1601.04839} {\path{arXiv:1601.04839}},
  \href {http://dx.doi.org/10.1103/PhysRevC.93.021902}
  {\path{doi:10.1103/PhysRevC.93.021902}}.

\bibitem{Nahrgang:2018afz}
M.~Nahrgang, M.~Bluhm, T.~Schäfer, S.~A. Bass, {Diffusive dynamics of critical
  fluctuations near the QCD critical point}\href
  {http://arxiv.org/abs/1804.05728} {\path{arXiv:1804.05728}}.

\bibitem{Cyrol:2017ewj}
A.~K. Cyrol, M.~Mitter, J.~M. Pawlowski, N.~Strodthoff, {Nonperturbative quark,
  gluon, and meson correlators of unquenched QCD}, Phys. Rev. D97~(5) (2018)
  054006.
\newblock \href {http://arxiv.org/abs/1706.06326} {\path{arXiv:1706.06326}},
  \href {http://dx.doi.org/10.1103/PhysRevD.97.054006}
  {\path{doi:10.1103/PhysRevD.97.054006}}.

\bibitem{Herbst:2010rf}
T.~K. Herbst, J.~M. Pawlowski, B.-J. Schaefer, {The phase structure of the
  Polyakov–quark–meson model beyond mean field}, Phys. Lett. B696 (2011)
  58--67.
\newblock \href {http://arxiv.org/abs/1008.0081} {\path{arXiv:1008.0081}},
  \href {http://dx.doi.org/10.1016/j.physletb.2010.12.003}
  {\path{doi:10.1016/j.physletb.2010.12.003}}.

\bibitem{Rennecke:2016tkm}
F.~Rennecke, B.-J. Schaefer, {Fluctuation-induced modifications of the phase
  structure in (2+1)-flavor QCD}, Phys. Rev. D96~(1) (2017) 016009.
\newblock \href {http://arxiv.org/abs/1610.08748} {\path{arXiv:1610.08748}},
  \href {http://dx.doi.org/10.1103/PhysRevD.96.016009}
  {\path{doi:10.1103/PhysRevD.96.016009}}.

\bibitem{Pawlowski:2017gxj}
J.~M. Pawlowski, N.~Strodthoff, N.~Wink, {Finite temperature spectral functions
  in the O(N)-model}\href {http://arxiv.org/abs/1711.07444}
  {\path{arXiv:1711.07444}}.

\bibitem{Cyrol:2018xeq}
A.~K. Cyrol, J.~M. Pawlowski, A.~Rothkopf, N.~Wink, {Reconstructing the
  gluon}\href {http://arxiv.org/abs/1804.00945} {\path{arXiv:1804.00945}}.

\bibitem{Floerchinger:2011sc}
S.~Floerchinger, {Analytic Continuation of Functional Renormalization Group
  Equations}, JHEP 05 (2012) 021.
\newblock \href {http://arxiv.org/abs/1112.4374} {\path{arXiv:1112.4374}},
  \href {http://dx.doi.org/10.1007/JHEP05(2012)021}
  {\path{doi:10.1007/JHEP05(2012)021}}.

\bibitem{Kamikado:2013sia}
K.~Kamikado, N.~Strodthoff, L.~von Smekal, J.~Wambach, {Real-time correlation
  functions in the $O(N)$ model from the functional renormalization group},
  Eur. Phys. J. C74~(3) (2014) 2806.
\newblock \href {http://arxiv.org/abs/1302.6199} {\path{arXiv:1302.6199}},
  \href {http://dx.doi.org/10.1140/epjc/s10052-014-2806-6}
  {\path{doi:10.1140/epjc/s10052-014-2806-6}}.

\bibitem{Tripolt:2013jra}
R.-A. Tripolt, N.~Strodthoff, L.~von Smekal, J.~Wambach, {Spectral Functions
  for the Quark-Meson Model Phase Diagram from the Functional Renormalization
  Group}, Phys. Rev. D89~(3) (2014) 034010.
\newblock \href {http://arxiv.org/abs/1311.0630} {\path{arXiv:1311.0630}},
  \href {http://dx.doi.org/10.1103/PhysRevD.89.034010}
  {\path{doi:10.1103/PhysRevD.89.034010}}.

\end{thebibliography}







\end{document}